\documentclass[useAMS,usenatbib,usegraphicx]{mn2e}

\usepackage{graphicx}
\usepackage{color, array}

\title[ANGULAR BROADENING: THE STRUCTURE FUNCTION]{Coronal turbulence and the angular broadening of radio sources - the role of the structure function}
\author[INGALE, SUBRAMANIAN \& CAIRNS]{M. Ingale$^{1}$\thanks{E-mail:
i.madhusudan@students.iiserpune.ac.in (MI)} 
P. Subramanian$^{1,2}$ and Iver Cairns$^{3}$\\
$^{1}$Indian Institute of Science, Education and Research, Pune, 411008, India\\
$^{2}$Center of Excellence in Space Sciences, India (http://www.cessi.in)\\
$^{3}$School of Physics, University of Sydney, Sydney, Australia\\}
\begin{document}


\pagerange{\pageref{firstpage}--\pageref{lastpage}} \pubyear{2002}

\maketitle

\label{firstpage}
 
\begin{abstract}
The amplitude of density turbulence in the extended solar corona, especially near the dissipation scale, impinges on several problems of current interest.
Radio sources observed through the turbulent solar wind are broadened due to refraction by and scattering off density inhomogeneities, and observations of scatter broadening are often employed to constrain the turbulence amplitude. The extent of such scatter broadening is usually computed using the structure function, which gives a measure of the spatial correlation measured by an interferometer. Most such treatments have employed analytical approximations to the structure function that are valid in the asymptotic limits $s \gg l_{i}$ or $s \ll l_{i}$, where $s$ is the interferometer spacing and $l_{i}$ is the inner scale of the density turbulence spectrum. We instead use a general structure function (GSF) that straddles these regimes, and quantify the errors introduced by the use of these approximations. We have included the effects of anisotropic scattering for distant cosmic sources viewed through the solar wind at small elongations. We show that the regimes where the GSF predictions are more accurate than those of the asymptotic expressions are not only of practical relevance, but are where inner scale effects influence estimates of scatter broadening. 
Taken together, we argue that the GSF should henceforth be used for scatter broadening calculations and estimates of turbulence amplitudes in the solar corona and solar wind.
\end{abstract}


\section{Introduction}

The nature of turbulence in the solar wind has been a subject of intense research. While considerable progress has been made, the nature of turbulent dissipation, especially in the extended solar corona, remains a significant unsolved problem. In particular, the damping of Alfv\'en turbulence on ions has attracted considerable recent interest following pioneering observations of perpendicular ion heating by the UVCS instrument aboard SOHO \citep {b23, b24, b25, b32, b16}. A major question in this regard is whether there is enough power in the turbulent cascade to enable direct perpendicular heating of ions \citep{b17}. Density fluctuations in the solar wind have the potential to constrain the power in low frequency kinetic Alfv\'enic turbulence, which in turn may be responsible for some heating of the corona \citep{b11} and acceleration of the solar wind. One way of addressing this question is via observations of the angular broadening of radio sources observed through the turbulent corona, which allows us to place constraints on the ``amplitude'' of the density turbulence spectrum, $C_{N}^{2}$. Recent observations, as well as analytical and numerical results e.g., \citet{b12} suggest that the fluctuations in the solar wind well below the proton scale are dominated by kinetic Alfv\'en turbulence. Consequently the magnetic field fluctuations and the density fluctuations are correlated, which implies that knowledge of $C_N^2$ can provide a good handle on estimating the power near the dissipation scale. The procedure involved in estimating $C_{N}^{2}$ from angular broadening observations employs the structure function that characterises the broadening. Treatments of angular broadening so far \citep{b14, b6, b39} have employed approximations to the structure function that are valid only for situations where the angular scales are either $\ll$ or $\gg$ than the inner (dissipation) scale. In this work, we have employed the general structure function (GSF) that does not use these approximations and straddles the $s \ll l_i$ and $s \gg l_i$ regimes. We demonstrate why the GSF should be used for accurate estimates of the extent of angular broadening for sources embedded in the solar corona as well as for distant celestial sources observed through the scattering screen of the solar wind. This study is also relevant to recent investigations of the turbulent properties of the ionised inter galactic medium (IGM) carried out using observations of the comapct radio sources \citep{b55, b56}.

The rest of this paper is organized as follows: in Section 2 we present a brief background of angular broadening and the relevance of the structure function. We introduce the asymptotic branches of the structure function and the GSF, treating anisotropic as well as isotropic scattering, using a density power spectrum which includes a power law and an exponential cutoff at the inner scale. The results for the difference between the predictions of the GSF and the asymptotic branches, together with the effect of local flattening of the turbulence spectrum and inner scale effects are presented in Section 3. Finally the results are interpreted and summarized in Section 4. 

\section[]{Angular Broadening of an ideal point source}
For the rest of this paper, we discuss the predictions of a given structure function regarding the angular broadening of an ideal point source. For distant celestial sources observed against the foreground of the solar wind, it is generally observed that the scatter-broadened images are highly anisotropic, especially for heliocentric distances $\leq 6 R_{\odot}$ \citep{b2, b20}. On the other hand, the scattering mechanism for sources embedded in the solar corona are generally isotropic, as will be discussed later. This motivates the need to develop a formalism that can handle anisotropic scattering, when required. We discuss this in more detail in \S~2.2

\subsection{Structure function}

A theoretical understanding of scatter broadening using wave optics in the paraxial approximation involves the phase structure function $D_{\phi}$. The phase structure function gives direct information on the extent to which an ideal point source is broadened. Since the phase structure function contains information about the spectrum of density turbulence, comparisons of the observed angular extent of sources with estimates using the phase structure function yields useful constraints on the quantities characterizing the turbulence spectrum. One such quantity is the amplitude of density turbulence ($C_N^2$), which is the normalizing constant of the spatial power spectrum of density fluctuations and is indicative of the power content in the spectrum at a given spatial scale.

Consider a Cartesian coordinate system with radiation propagating in the +z direction. 
Suppose the turbulent medium fills the half space $z > 0$; $x$ and $y$ represent transverse coordinates. 
A coherent wave incident on the turbulent medium at $z = 0$ experiences loss of spatial as well as 
temporal coherence due to refraction by and scattering off the density inhomogeneities associated with 
the turbulence. An interferometer measures the mutual coherence function at z = L. e.g., \citet{b6}

\begin{equation}
\Gamma(\mathbf{s}) = \frac{\langle E(\mathbf{r})E^{*}(\mathbf{r_1})\rangle}{\langle |E|^{2}\rangle} \,\, ,
\label{mutcoh}
\end{equation}

where $\mathbf{r} = (x, y; z = L)$ and $\mathbf{r_1} = \mathbf{r} + \mathbf{s}$, with $\mathbf{s} = (x', y', 0)$ are the transverse coordinates. In the context of an interferometer on the Earth measuring the correlation expressed by (\ref{mutcoh}), the quantity $|\mathbf{s}|$ can be interpreted as the interferometer baseline; i.e., the separation between two antennas transverse to the line of sight. For small angle scattering the mutual coherence function $\Gamma(\mathbf{s})$ is related to the phase structure function $D_{\phi}(s)$ through (Ishimaru 1978, Ch 20)

\begin{equation}
\Gamma(\mathbf{s}) = \exp(-D_{\phi}(\mathbf{s})/2) \,\, .
\label{visibility}
\end{equation}

\noindent Alternatively the phase structure function is also defined by \citep{b15},

\begin{eqnarray}
D_{\phi}(s) &=& \langle[\phi(\mathbf{r})-\phi(\mathbf{r}+\mathbf{s})]^2\rangle] \nonumber \\
      &=& 2[\langle\phi(\mathbf{r})^2\rangle - 
        \langle\phi(\mathbf{r})\phi(\mathbf{r}+\mathbf{s})\rangle] \,\, ,
\label{def}
\end{eqnarray}

\noindent where $\phi(\mathbf{r})$ is the phase deviation calculated along the line of sight through a turbulent medium. The value of $\mathbf{s}$ where $D_{\phi}(\mathbf{s}) = 1$ gives a measure of the extent to which an ideal point source is broadened due to the effects of scattering on turbulent density fluctuations.
The relation between the phase fluctuations $\delta \phi(\mathbf{r})$ experienced by a wave propagating through the turbulent medium and the density fluctuations $\delta N_e(\mathbf{r})$ is particularly simple for the thin screen geometry, where the density irregularities are assumed to be concentrated across a two dimensional scattering screen of thickness $\Delta L$. When $\Delta L \ll$ the distance between the source and the observer ($L$), the change in the phase of the wave can be expressed as \citep{b6}

\begin{equation}
\delta \phi(\mathbf{r}) = r_e \lambda \delta N_e(\mathbf{r}) \Delta L \,\, .
\end{equation}

\noindent Here $r_e$ is the classical electron radius and $\lambda = 2\pi c/f$ is the radiation wavelength. With this the phase structure function defined in (\ref{def}) can be written in terms of density fluctuations as :

\begin{equation}
D_{\phi}(\mathbf{s}) = 4 \pi r_e^2 \lambda^2 [\langle \delta N_e(\mathbf{r})^2 \rangle -
               \langle \delta N_e(\mathbf{r})\delta N_e(\mathbf{r}+\mathbf{s})\rangle]
               \Delta L \,\, .
\end{equation}

\noindent Here $\langle\delta N_e(\mathbf{r})\delta N_e(\mathbf{r}+\mathbf{s})\rangle$ is the spatial correlation function of the electron density fluctuations. The Fourier transform of the spatial correlation function yields the spatial power spectrum

\begin{equation}
\langle\delta N_e(\mathbf{r})\delta N_e(\mathbf{r}+\mathbf{s})\rangle = \int_{-\infty}^{\infty}\! S_n(\mathbf{k},R)\mathrm{exp}
(i\mathbf{k\cdot s}) \, \mathrm{d^2}k \,\, .
\label{fourier}
\end{equation}

\noindent Here $\mathbf{k} = (k_x, k_y, k_z=0)$ is the transverse wavenumber and $S_n(k, R)$ is the spatial power spectrum of the fluctuating part of the electron density which depends upon the wavenumber $k$ as well as the heliocentric distance $R$.

In order to proceed further, we define a cartesian coordinate 
system $x, y, z$, with $z$ along the line of sight. 
For plane wave propagation, which is relevant for the case of radiation from distant background celestial sources viewed against the turbulent solar wind scattering screen, density inhomogeneities are concentrated 
in a thin screen of thickness $\Delta L$, located at 
$z = 0$ between the source and the observer. In this case, the transverse coordinates $x$ and $y$ are in the plane of the scattering screen, 
which is perpendicular to the line of sight. 
The $x$ coordinate is taken to be along the projection of the local magnetic field vector into the $x-y$ plane and at small elongations, it is observed that scatter broadened images are typically stretched along the $x$ direction. As explained later, this is treated using a formalism where the underlying turbulent eddies are also elongated in the $x$ direction.

In case of spherical wave propagation, which is relevant for observations of sources embedded in the solar corona, the line of sight (z-axis) is along the radial 
direction. The line of sight is substantially aligned with the magnetic field. As we will see later, the scattering process in this case is isotropic, and the transverse coordinates $x$ and $y$ are treated on an equal footing.

Observations of solar wind density fluctuations in the low speed solar wind between 0.3 and 1 AU at frequencies $< 0.1$ Hz (which is generally accepted to be in the inertial range) reveal that the spatial power spectrum of electron density fluctuations in the solar corona $S_{n}(k)$ largely follow the isotropic Kolmogorov scaling \citep{b52, b41, b15}. In-situ observations near the Earth at higher frequencies find no evidence for any deviation from the Kolmogorov scaling as a function of the angle with respect to the local magnetic field direction \citep{b53, b41}.
At higher frequencies, however, there is evidence for steepening at wave numbers $\approx 2\pi/l_i$. In high speed solar wind streams, there seems to be some evidence for spectral flattening at high frequencies prior to the inner scale. The quantity $l_{i}$ is generally referred to as the inner scale, where dissipation sets in. The steepening of the spectrum beyond the inner scale is often attributed to the dissipation of the turbulent eddies and associated waves \citep{b13, b45}. Observations of the radial dependence of the inner scale have been made by \citet{b46}, \citet{b38}, \citet{b15}, \citet{b2} and \citet{b28}. One model for the dissipation mechanism is cyclotron damping of hydromagnetic waves \citep{b15, b47}. However, observations by \citet{b5}, \citet{b1} and \cite{b34} suggest that the turbulent spectra might also exhibit breakpoints near the proton and electron gyroscales.

\subsubsection{Anisotropic scattering}

As mentioned earlier, scatter-broadened images of distant celestial sources viewed on foreground of the solar wind at small elongations from the Sun tend to be strongly anisotropic \citep{b2, b15, b14}. The scatter-broadened images are observed to be elongated along the direction of the (predominantly radial) large-scale magnetic field. One must therefore consider the effect of anisotropy while calculating the phase structure function in these cases. The thin screen geometry is found to be appropriate for this situation 
\citep{b14, b15}.

The turbulent density spectrum is commonly modeled as a power law in wavenumber space. It is known that at the smallest scales the density spectrum displays abrupt steepening \citep{b15} indicating the existence of the inner scale. In this region an exponential cut-off is often a good approximation to a steeper power law \citep{b5, b1}. Models also suggest that the dissipation range is an exponential cutoff, implying that observations of steeper power laws might arise from instrumental limitations \citep{b21}.

\citet{b49} developed a theory of refractive scintillation that includes anisotropy. They model the turbulent spectrum as a power law with the power law index $\alpha \leq 4$, but do not include a cutoff due to dissipation. We extend the formulation of \citet{b49} to include the exponential cut-off together with the power law spectrum. The anisotropic generalization of the power law spectrum with exponential cutoff leads to the following form for the power spectrum of density fluctuations:

\begin{eqnarray}
	\lefteqn {S_n(\mathbf{k},R) } \nonumber \\
		&=& C_N^2(R) \, (\rho^2 k_x^2 + k_y^2)^{-\alpha/2}
        	\exp[-(\rho^2 k_x^2 + k_y^2)(\mathbf{l_{i}}/2 \pi)^{2}] .
\label{anispectrum}
\end{eqnarray}

\noindent The quantity $C_N^2$ is the amplitude of the density turbulence \citep{b33}, 
$\alpha$ is the power law index and the exponential turnover occurs at an inner scale $\mathbf{l_i}$. The parameter $\rho$ (which is $> 1$) measures the degree of anisotropy, which can be interpreted in the following way : if the density blob in the screen has length $l$ in the y direction, it is elongated to $\rho l$ in the x direction. In writing (\ref{anispectrum}) we have assumed for the sake of simplicity that $\rho$ is independent of the spatial scale following \citet{b49}. The inner scale ($\mathbf{k_i} = 2\pi/\mathbf{l_i}$) is also assumed to be anisotropic and follows the same scaling as $\mathbf{k}$.

Anisotropy can be treated by replacing the usual Cartesian coordinate system with coordinates $k_r$ and $\xi$ defined as

\begin{equation}
k_r = (\rho^2 k_x^2 + k_y^2)^{1/2} \,\, ; \,\,\,
\xi = tan^{-1}\left(\frac{k_y}{\rho k_x}\right) \,\, .
\end{equation}

Accordingly the area element will be scaled with the Jacobian, $|J| = k_r/\rho$.

\begin{equation}
d^2k = \frac{k_r}{\rho} dk_rd\xi \,\, .
\end{equation}

\noindent Using equation~(\ref{fourier}) in (\ref{anispectrum}) gives

\begin{eqnarray}
   \lefteqn{\langle\delta N_e(\mathbf{r})\delta N_e(\mathbf{r}+\mathbf{s})\rangle 
     = \frac{1}{\rho} 8\pi^2 r_e^2 \lambda^2 \Delta L C_N^2(R)}  \nonumber \\
     & & \times \int_{0}^{\infty} J_0(k_r s) k_r^{1-\alpha} \exp[-(k_r l_i/2)^2] \, \mathrm{d}k_r \,\, ,
\label{dencorr}
\end{eqnarray}

\noindent where $J_0$ is the Bessel function of the first kind. From the definition of the phase structure function, we can write,

\begin{eqnarray}
\lefteqn {D_{\phi}(\mathbf{s}) = \frac{1}{\rho} 8\pi^2 r_e^2 \lambda^2 \Delta L C_N^2(R)} 
\nonumber \\
   & & \times \int_{0}^{\infty} [1-J_0(k_rs)] k_r^{1-\alpha} \exp[-(k_r l_i/2)^2] \, 
   \mathrm{d}k_r \,\, .
  \label{pridphi}
\end{eqnarray}

\vskip1ex
\noindent On integrating over $k_r$ we obtain

\begin{eqnarray}
D_{\phi}(\mathbf{s}) = \frac{1}{\rho}\frac{8\pi^2r_e^2\lambda^2 \Delta L}{2^{\alpha-2}(\alpha-2)}
\Gamma\left(1-\frac{\alpha-2}{2}\right) C_N^2(R) l_i^{\alpha-2}(R) \nonumber \\
   \times\left\lbrace_1F_1\left[-\frac{\alpha-2}{2},1,-\left(\frac{s}{l_i(R)}\right)^2
 \right]-1  \right\rbrace \  \,\, .
\label{eqcsfani0}
\end{eqnarray}

\noindent Including the effects of the spatially varying plasma frequency $f_p(R)$ \citep{b8} gives
\begin{eqnarray}
D_{\phi}(\mathbf{s}) = \frac{1}{\rho} \frac{8\pi^2r_e^2\lambda^2 \Delta L}{2^{\alpha-2}(\alpha-2)} \Gamma\left(1-\frac{\alpha-2}{2}\right) \frac{C_N^2(R) l_i^{\alpha-2}(R)}
{(1-f_p^2(R)/f^2)} \nonumber \\
   \times\left\lbrace_1F_1\left[-\frac{\alpha-2}{2},1,-\left(\frac{s}{l_i(R)}\right)^2
 \right]-1  \right\rbrace \ \, .
\label{eqcsfani}
\end{eqnarray}

\noindent Here $_1F_1$ denotes the confluent hypergeometric function. For $2 < \alpha < 4$, we find the following limiting forms of equation (\ref{eqcsfani}) for anisotropic scattering:

\begin{equation}
D_{\phi}(\mathbf{s}) = \frac{1}{\rho} \frac{4\pi^2 r_e^2 \lambda^2 \Delta L}
				{2^{\alpha-2}} \,\, \Gamma\left(1-\frac{\alpha-2}{2}\right)
				\frac{C_N^2(R) l_i^{\alpha-4}(R)}{(1-f_p^2(R)/f^2)}s^2 \, \, ,
\label{eqslliani}
\end{equation} 

\noindent for $s \ll l_i$, and 

\begin{equation}
D_{\phi}(\mathbf{s}) = \frac{1}{\rho} \frac{8 \pi^2 r_e^2 \lambda^2 \Delta L}
		{2^{\alpha-2}(\alpha-2)} \,\, \frac{\Gamma\left(1-(\alpha-2)/2\right)}{\Gamma\left(1+(\alpha-2)/2\right)} \frac{C_N^2(R) s^{\alpha-2}}{(1-f_p^2(R)/f^2)} \,\, ,
 	\label{eqsgliani}
\end{equation}

\noindent for $s \gg l_i$. In the preceding equations, $ |\mathbf{s}| = (\frac{s_x^2}{\rho^2} + s_y^2)^{\frac{1}{2}}$, and $l_i = (\frac{l_{ix}^2}{\rho^2} + l_{iy}^2)^{\frac{1}{2}}$. \\

\noindent The anisotropic coherence length is defined as

\begin{equation}
D_{\phi}(\mathbf{s_0}) = D_{\phi}(s_{0x}, s_{0y}) = 1 \,\, .
\label{root}
\end{equation}

We note that the root of (\ref{eqcsfani}) is $|\mathbf{s_0}| = (s_{0x}^2 + s_{0y}^2)^{\frac{1}{2}}$. Following \citet{b49}, we assume that the coherence length in x direction $s_{0x}$ is elongated by the factor of $\rho$ relative to the coherence length $s_{0y}$ in the y direction; in other words, 

\begin{equation}
s_{0x} = \rho s_{0y}
\label{elongate}
\end{equation}

Psuedo-codes for implementing the GSF and numerically evaluating the coherence length are given in the appendix. We assume a similar relation for the inner scale: $l_{ix} = \rho l_{iy}$. 
Using (\ref{root}) and (\ref{elongate}) we obtain the following expressions for the semi-major axes of the scatter-broadened image projected on the scattering screen in terms of the coherence length $s_{0x}$ and $s_{0y}$ :

\begin{equation}
\theta_{cx} = (2\pi s_{0x}/\lambda)^{-1} \, \, ,
\label{anianglex}
\end{equation}

\begin{equation}
\theta_{cy} = (2\pi s_{0y}/\lambda)^{-1} = \rho \theta_{cx} \,\, .
\label{aniangley}
\end{equation}

Note that expressions (\ref{pridphi}) $-$ (\ref{eqsgliani}) depend on $\rho$ only through a factor $1/\rho$, while the scattering angles (\ref{anianglex}) and (\ref{aniangley}) in the $x$ and $y$ direction respectively differ only by a factor of $\rho$. With isotropic results recovered in the case $\rho = 1$, this suggests that anisotropic scattering is relatively simple to incorporate.

\subsubsection{Isotropic scattering} 

There are situations where isotropic scattering is a reasonable assumption. 
If the underlying turbulent eddies are isotropic, this is obviously justified. However, it is widely assumed that this is not the case; i.e., the underlying turbulent spectrum is in fact anisotropic. However, the extent of anisotropy observed in scatter-broadened images is determined more by the variation in the direction of the large scale magnetic field with respect to the line of sight than by the degree of anisotropy of the density fluctuations \citep{b10}. Using an anisotropic Goldreich-Sridhar spectrum, these authors have shown that the scatter broadened images will be isotropic if the direction of the large-scale magnetic field is substantially aligned with the line of sight. This is intuitively obvious, since the plasma response would be gyrotropic about the large-scale magnetic field. Specifically, if $\gamma$ is the angle between the magnetic field and the line of sight, they show that if $\gamma \ll (s/l_{out})^{1/3}$, where s is the baseline and $l_{out}$ is the outer scale of the turbulence, the dominant contribution to the turbulent spectrum comes from the values of $k_x$ and $k_y$ satisfying $k_x^2 + k_y^2 \simeq s^{-2}$, and $S_n(k,R)$ is nearly isotropic.

\indent In the case of spherical wave propagation, where the observer is looking through the corona down at a source on the Sun, the line of sight from the Earth to the Sun is radial, and $\gamma$ satisfies this condition amply.
In this situation the effects of anisotropic scattering are likely to be minimal. Images of scatter-broadened sources {\bf near disk center} in the solar corona are indeed not very anisotropic \citep{b48, b29}, validating this idea. Stronger support for this argument is provided by the fact that type I radio bursts are strongly circularly polarised near disk center, and become less so near the limb. Since quasi-transverse magnetic field regions (in this case, regions of horizontal magnetic field) serve as depolarization sites, this implies that sources that are substantially near disk center do not encounter such regions, at least above the level where the emission originates; in other words, the magnetic field along the line of sight is largely radial \citep{b54}. As discussed above, this implies that the scattering process will be isotropic. There is also some evidence for the fact that the inner scale itself is isotropic, in which case the assumption of isotropy is well justified for scales comparable to or less than the inner scale \citep{b4, b6}.
For an isotropic turbulent density spectrum with an inner scale $l_i$, (\ref{anispectrum}) reduces to 

\begin{equation}
S_n(k,R) = C_N^2(R) \, k^{-\alpha} \exp[-(k l_i/2 \pi)^2] \,\, ,
\label{spectrum}
\end{equation}

\noindent where the spatial wavenumber $k = (k_x^2 + k_y^2)^{\frac{1}{2}}$.

Therefore the phase structure function, (\ref{eqcsfani}) can be rewritten as \citep{b15},

\begin{eqnarray}
D_{\phi}(s) = \frac{8\pi^2r_e^2\lambda^2 \Delta L}{2^{\alpha-2}(\alpha-2)}
\Gamma\left(1-\frac{\alpha-2}{2}\right) \frac{C_N^2(R) l_i(R)^{\alpha-2}}
{(1-f_p^2(R)/f^2)} \nonumber \\
   \times\left\lbrace_1F_1\left[-\frac{\alpha-2}{2},1,-\left(\frac{s}{l_i(R)}\right)^2
 \right]-1  \right\rbrace \,\, ,
\label{eqcsf}
\end{eqnarray}

\noindent and the corresponding asymptotic branches (\ref{eqslliani}) and (\ref{eqsgliani}) take the form \citep{b14, b39}:

\begin{equation}
D_{\phi}(s) = \frac{4\pi^2 r_e^2 \lambda^2 \Delta L}{2^{\alpha-2}}
\Gamma\left(1-\frac{\alpha-2}{2}\right)s^2 \frac{C_N^2(R)l_i(R)^{\alpha-4}}{(1-f_p^2(R)/f^2)} \,\, ,
\label{eqslli}
\end{equation} 

\noindent for $s \ll l_i$ and for $s \gg l_i$, 
\begin{equation}
D_{\phi}(s) = \frac{8 \pi^2 r_e^2 \lambda^2 \Delta L}{2^{\alpha-2}(\alpha-2)}
\frac{\Gamma\left(1-(\alpha-2)/2\right)}{\Gamma\left(1+(\alpha-2)/2\right)}  
\frac{C_N^2(R) s^{\alpha-2}}{(1-f_p^2(R)/f^2)} \,\, .
\label{eqsgli}
\end{equation}

\noindent The isotropic coherence length $s_0$ is defined by 

\begin{equation}
D_{\phi}(s_0) = 1 \,\, .
\label{eqcohlength}
\end{equation}

\noindent For a given wavelength $\lambda$, the extent to which an ideal 
point source is broadened is given in terms of the isotropic coherence length $s_{0}$ as

\begin{equation} 
\theta_c = (2\pi s_0/\lambda)^{-1} \,\, .
\label{eqthetac}
\end{equation}

Just as the anisotropic coherence length $\mathbf{s}_0 = (s_{0x}, s_{0y})$ can be readily calculated from the asymptotically correct expressions (\ref{eqslliani}) and (\ref{eqsgliani}) for the structure function (\ref{eqcsfani}), the isotropic coherence length $s_0$ can be calculated easily for the asymptotic approximations (\ref{eqslli}) and (\ref{eqsgli}) of the phase structure function (\ref{eqcsf}). Several authors e.g., \citep{b6, b39} have used these asymptotic expressions to obtain estimates of angular broadening of sources in the solar corona. However, there are limitations associated with using these asymptotic expressions. Specifically, equations (\ref{eqslliani}) and (\ref{eqsgliani}) or (\ref{eqslli}) and (\ref{eqsgli}) do not meet seamlessly at $s = l_i$, which suggests that in situations where the baseline is comparable to the inner scale, the asymptotic approximations can not give reliable results \citep{b14, b39}.

In what follows, we compute the scatter broadening angles $\theta_{cx}$ and $\theta_{cy}$ for the anisotropic case using the full expressions (\ref{eqcsfani}) and compare them with those obtained with the limiting expressions (\ref{eqslliani}) and (\ref{eqsgliani}).Similarly, for the isotropic case we compare scatter broadening angles $\theta_c$ obtained using full expression (\ref{eqcsf}) with those obtained by using the limiting expressions (\ref{eqslli}) and (\ref{eqsgli}).

\subsection{Spherical versus plane wave propagation}
The extent of scatter broadening depends on whether the wavefront is planar (1‐D) or spherical (3‐D). When a source is embedded in the scattering medium, as is the case for sources in the solar corona \citep{b6, b39} it is appropriate to adopt a formalism that includes the spherically diverging nature of the wavefront. As discussed earlier (\S~2.1.2), the assumption of isotropy is also justified in this situation.
For the spherically diverging wavefront, the observer is sensitive to a range of eddy sizes given by $s a/b$, where $s$ is the interferometer baseline, $a$ is the distance of the scattering screen from the source and $b$ is the distance of the observer from the source; see \citet{b39} for details. In other words, the effective baseline for spherical wave propagation at a given heliocentric distance $R$ is \citep{b22}

\begin{equation}
s_{\rm eff} = sR/(R_1-R_0)\, ,
\label{seff}
\end{equation}

\noindent where $R_{1}$ is the heliocentric distance at which the observation of angular broadening is made and $R_{0}$ is the heliocentric distance at which the source is situated.

The planar (1D) formalism, on the other hand, is appropriate when the source, scattering region(s), and observer are all far apart from each other, as assumed in calculations for pulsars and other celestial sources viewed through the scattering screen of the solar wind. As discussed earlier (\S~2.1.1), a treatment of anisotropic scattering is essential for this situation. In this case an observer is typically sensitive only to scattering regions (eddies) with sizes of the order of the interferometer baseline $s$. In other words, $s_{\rm eff} = s$ for plane wave propagation.

\subsection{Density Models}

To estimate the angular broadening we need to integrate the random phase fluctuations along the line of sight. The lower limit of this integral is the fundamental plasma level ($f = f_p$) and so depends on the observing frequency, $f$. A model for $f_{p}$ is needed to fully describe refractive index and inner scale effects. Since $f_{p}^2 \propto n_{e}$ a model for the ambient electron density $n_{e}(R)$ is required.


One density model we use is due to \citet{b9}, which we will call the ``wind-like'' density model from now on. The electron density as a function of heliocentric distance $R$ (which is measured in units of $R_{\odot}$) is given by:
\begin{equation}
n_e(R) = 1.58 \times 10^{27} \times (R - 1)^{-2}\,\,\,   {\rm cm}^{-3} \,\, .
\label{wind}
\end{equation}

\noindent In fact, the wind-like density model only specifies that the density should be proportional to $(R - 1)^{-2}$; we obtain the proportionality constant of $1.58 \times 10^{27} \,\, {\rm cm^{-3}}$ by demanding that the density predicted by this model equal that predicted by the \citet{b50} density model at 1 AU. We also use the commonly used 4-fold Newkirk density model in this paper, which has a different proportionality constant (equal to ${\rm 4.2 \times 10^4 \,\, cm^{-3}}$).

\subsection{Amplitude of Density Turbulence : $C_N^2(R)$}

In order to characterize the amplitude $C_{N}^{2}$ of the turbulent density spectrum $S_{n}$ (eq~\ref{spectrum}), we use a model originally proposed by \citet{b3} and later refined by 
\citet{b37}. This model is based on VLBI observations in the outer corona and solar wind. 
\citet{b37} obtained the following expression for $C_N^2$ as a linear fit to scattering data between $10 R_{\odot} - 50 R_{\odot}$:

\begin{equation}
C_N^2 = 1.8 \times 10^{10}\left(\frac{R}{10R_{\odot}}\right)^{-3.66}  {\rm m}^{-20/3}\,\, .
\label{ampltd}
\end{equation}

\noindent The dimensions of $C_N^2(R)$ in general are m$^{-\alpha-3}$, where $\alpha$ is the power law index characterizing the inertial range of the turbulent density spectrum $S_{n}$ in (\ref{spectrum}).  It may be noted that (\ref{ampltd}) is valid only for a Kolmogorov spectrum ($\alpha = 11/3$). There is considerable evidence supporting the idea that inertial range density fluctuations in the solar wind follow the  Kolmogorov scaling \citep{b30, b19, b27, b35,b52, b41, b15}. 

While we primarily use the Kolmogorov scaling ($\alpha = 11/3$) in this work, there is some evidence for flattening of the spectrum (at heliocentric distances of a few $R_{\odot}$) from scales around 100 km down to the inner scale  \citep{b15, b6}, which might be evidence for the dispersion range that occurs prior to the dissipation range. The extent of this flattening strongly depends upon the phase of the solar cycle and the speed of the solar wind \citep{b28}. The evolution of the flattening feature with heliocentric distance is not known.

We have examined this issue by using $\alpha = 3$ in (\ref{spectrum}), as in \citet{b6}. 
Since (\ref{ampltd}) as it stands is valid only for $\alpha = 11/3$, we need to re-calculate $C_N^2$ using equation (9) of \citet{b37} with $\alpha = 3$. We obtained a least square fit to the plot of the newly calculated $C_N^2$ against the impact parameter $R_0$, which yielded the following modified model for $C_N^2$ for $\alpha = 3$:
\begin{equation}
C_N^2 = 8.1 \times 10^{12}\left(\frac{R}{10R_{\odot}}\right)^{-3.66}  {\rm m}^{-6} \,\, .
\label{newcnsq}
\end{equation}

\noindent Although we have primarily used (\ref{ampltd}), which holds for $\alpha = 11/3$, we also discuss the modifications to our results arising from the use of (\ref{newcnsq}) in \S~4 below.

\section{Results}

\subsection{When is the GSF needed?}

We start by exploring the circumstances under which the GSF, given by (\ref{eqcsfani}) for anisotropic scattering and (\ref{eqcsf}) for isotropic scattering, is significantly more accurate than the asymptotic branches (given by (\ref{eqslliani}) and (\ref{eqsgliani}) for anisotropic scattering and (\ref{eqslli}) and (\ref{eqsgli}) for isotropic 
scattering). We will use the coherence length $\mathbf{s_{0}}$ in order to address this question. For anisotropic scattering we recall that the coherence length $\mathbf{s_{0}} = (s_{0x}, s_{0y})$ is related to the phase structure function via (\ref{root}), and the scattering angles $\theta_{cx}$ and $\theta_{cy}$ (which would be the semi-major axes of an image corresponding to an ideal point source subject to scatter broadening) are related to $s_{0x}$ and $s_{0y}$ via (\ref{anianglex}) and (\ref{aniangley}), respectively. In what follows, we use the GSF defined in (\ref{eqcsfani}) and the asymptotic branches in (\ref{eqslliani}) and (\ref{eqsgliani}) to calculate $(s_{0x}, s_{0y})$ and $(\theta_{cx},\theta_{cy})$. In order to avoid confusion we discuss only the y-component of the coherence length, namely $s_{0y}$. The results for the x-component of the coherence length, $s_{0x}(= \rho s_{0y})$ are identical.

In order to compare the angular broadening predictions of the GSF with those predicted by the asymptotic branches, we plot the relative error introduced in the coherence length $s_{0y}$ when either of the asymptotic approximations (\ref{eqslliani}) or (\ref{eqsgliani}) of the GSF (\ref{eqcsfani}) is used, as a function of the inner scale $l_{iy}$. Since we work with the relative error in $s_{0y}$, our conclusions are independent of the observing frequency for plane wave propagation. We carry out a similar exercise for isotropic scattering, relevant for the spherically diverging wavefront. In this situation, as we
will see below in \S~3.1.2, (\ref{eqcsf}) is modified to include an integral along the line of sight. The lower limit of the line-of-sight integral, which is the plasma level, is a function of an observing frequency. The spherical wave calculations are therefore expected to be sensitive to the observing frequency. When the relative error is significant the predictions of the GSF disagree with those of the asymptotic branches, and the converse is true when the relative error is negligible. The inner scale $l_{i}$ is maintained as a free parameter in sections 3.2 -- 3.4 . 
When using the GSF, we note that the coherence length $\mathbf{s_{0}}$ needs to be calculated using a numerical root finding procedure. On the other hand, one can obtain an explicit analytical expression for $\mathbf{s_{0}}$ when using the asymptotic branches.
For non-zero, spatially varying ratios $f_p(z)/f$, the expressions for the GSF and the asymptotic branches for an anisotropic case are modified by the inclusion of the factor $[1-f_p^2(z)/f^2]^{-1}$ in (\ref{eqcsfani}), (\ref{eqsgliani}) and (\ref{eqslliani}). It can be easily shown that (\ref{eqcsfani0}) is recovered in the limit $f_p(z)/f \rightarrow 0$ (i.e.$f_p(z) \ll f$) and constant $f_p/f$). Similarly for an isotropic case the GSF and the asymptotic branches are modified to ((\ref{eqcsf}), (\ref{eqslli}) and (\ref{eqsgli})). Consequently, in the limit $f_p(z)/f \to 0$ the angular broadening results are independent of the ratio $f_p/f$ [Cairns, 1998].

\subsubsection{Plane wave propagation}

Plane wave propagation is relevant to the situation where one is observing a distant cosmic source against the background of the solar wind. In this situation, anisotropic scattering is important, especially for sources at small solar elongations. In what follows we compute the coherence length by using the anisotropic phase structure function discussed in \S~2.1.1, and compare it with those obtained by using the asymptotic approximations.

\begin{figure}
	\includegraphics[width=84mm]{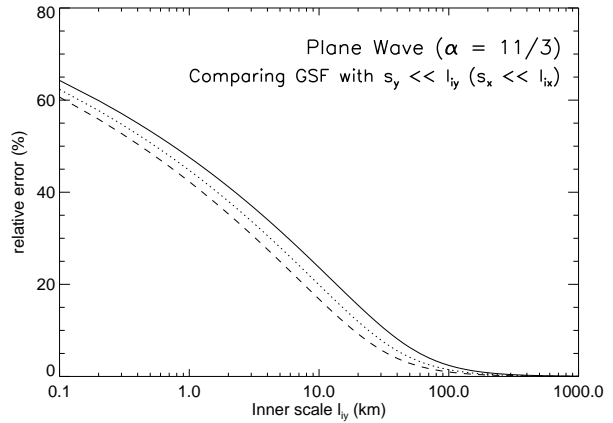}
	\caption{Relative error in the coherence length $s_{0y} \,\, (s_{0x})$ as a function of $l_{iy} \,\, (l_{ix})$ when the asymptotic branch $s_y \ll l_{iy} \,\, (s_x \ll l_{ix})$, (\ref{eqslliani}) is used. The calculations are for plane wave propagation through the corona and solar wind and for a representative solar elongation of $10R_{\odot}$. The solid line uses the degree of anisotropy, $\rho = 1$, the dotted line is for $\rho = 5$ and the dashed line uses, $\rho = 10$.}
	\label{plnrat1}
\end{figure}

\begin{figure}
	\includegraphics[width=84mm]{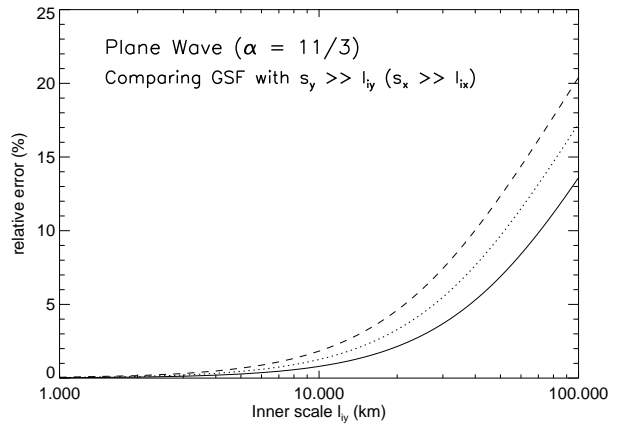}
	\caption{Relative error in the coherence length $s_{0y} \,\, (s_{0x})$ as a function of $l_{iy} \,\, (l_{ix})$ when the asymptotic branch $s_y \gg l_{iy} \,\, (s_x \gg l_{ix})$, (\ref{eqsgliani}) is used. The calculations are for plane wave propagation through the corona and solar wind and for a representative solar elongation of $10R_{\odot}$. The solid line uses the degree of anisotropy, $\rho = 1$, the dotted line is for $\rho = 5$ and the dashed line uses, $\rho = 10$.}
	\label{plnrat2}
\end{figure}

The coherence length $s_{0y}$ is calculated by using (\ref{root}) via (\ref{elongate}). Figures~\ref{plnrat1} and \ref{plnrat2} show the relative error in the predictions of the  $s_{0y}$ as a function of the inner scale $l_{iy}$, when either of the asymptotic branches is favoured over the GSF. We obtained $s_{0y}$ as a root of the GSF, (\ref{eqcsfani}) and compared with those of the asymptotic branches, (\ref{eqslliani}) and (\ref{eqsgliani}) for three different cases of the degree of anisotropy, $\rho$. The solid line is for $\rho = 1$, dotted line for $\rho = 5$ and the dashed line for $\rho = 10$. The results shown in Figures~\ref{plnrat1} and ~\ref{plnrat2} are for a representative solar elongation of $10R_{\odot}$. It is clear that anisotropy effects are not very significant; varying $\rho$ by a factor of 10 results in a difference of $< 10\%$ in the relative error.

Figure~\ref{plnrat1} shows that the region where the relative error is significant decreases with increasing degree of anisotropy. For $\rho = 1$, (solid line) the relative error increases sharply for $l_{iy} \leq 200$km; in this region, the $s_y \ll l_{iy}$ branch is inadequate and the GSF should be used. For $\rho = 5$, (dotted line) the relative error becomes significant for $l_{iy} \leq 100$km and for $\rho = 10$, (dashed line), the relative error is significant for $l_{iy} \leq 80$km. 

Figure~\ref{plnrat2} displays the corresponding relative error for the asymptotic branch $s_y \gg l_{iy}$. It is clear from the Figure that the extent of the region for which the relative error is significant increases with the degree of anisotropy. For $\rho = 1$, (solid line) the relative error is significant for $l_{iy} \geq 10$km, implying the asymptotic branch $s_y \gg l_{iy}$ is inadequate and the GSF should be used. For $\rho = 5$, the relative error increases for $l_{iy} \geq 8$km and for $\rho = 10$, the relative error is significant for $l_{iy} \geq 4$km.

From Figures \ref{plnrat1} and \ref{plnrat2} it is clear that, for plane wave propagation through the solar wind and corona, the coherence length $s_{0y}$ (and therefore the broadening angle computed from it) computed via the GSF agrees with the asymptotically correct expressions (i.e. relative error is negligible) only outside the range 4 km $\leq l_{iy} \leq$ 200 km. In other words, for the degree of anisotropy ranging from 1$-$10, the statement $s_y \ll l_{iy}$ is valid for $l_{iy} \geq 200$ km, and the predictions of (\ref{eqslliani}) hold well. Similarly, $s_y \gg l_{iy}$ is valid for $l_{iy} \leq 4$km, and the predictions of (\ref{eqsgliani}) will be accurate for values of the inner scale satisfying this condition. At a solar elongation of 10 $R_{\odot}$, we find that the GSF predictions disagree with those of the asymptotic branches for 4km $\leq l_{iy} \leq$ 200 km. Although the results shown in Figures \ref{plnrat1} and \ref{plnrat2} hold only for a solar elongation of 10 $R_{\odot}$, we have also investigated this aspect for other elongations. We find that the $s_y \ll l_{iy}$ asymptotic branch is valid for elongations $< 5 R_{\odot}$, while the $s_y \gg l_{iy}$ asymptotic branch is valid for elongations $> 20 R_{\odot}$. Thus for elongations between 5 and 20 $R_{\odot}$, the GSF needs to be used.

\subsubsection{Spherical wave propagation}

\begin{figure}
	\includegraphics[width=84mm]{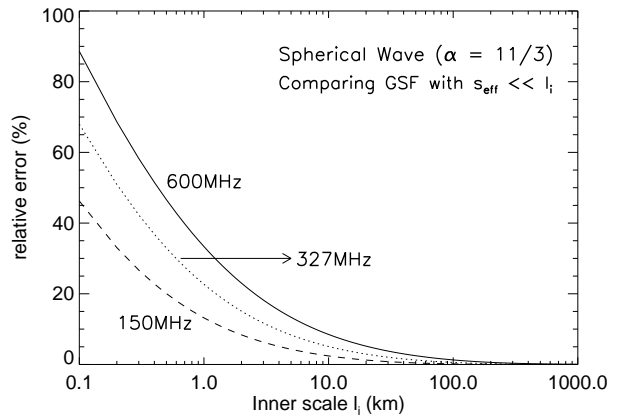}
	\caption{Relative error in the coherence length $s_{0}$ as a function of $l_i$ when 
the asymptotic branch $s_{\rm eff} \ll l_i$, (\ref{eqslli}) is used. The calculations are 
for spherical wave propagation appropriate for sources embedded in the corona. The solid 
line uses an observing frequency, $f = 600$MHz, the dotted line uses $f = 327$MHz and the 
dashed line uses $f = 150$MHz.}
	\label{sphrat1}
\end{figure}

\begin{figure}
	\includegraphics[width=84mm]{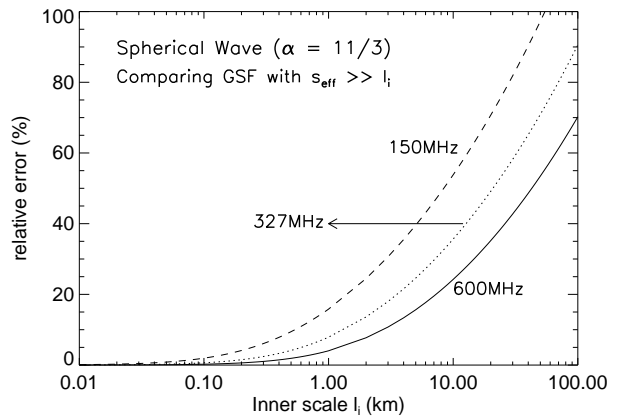}
	\caption{Relative error in the coherence length $s_{0}$ as a function of $l_i$ when 
the asymptotic branch $s_{\rm eff} \gg l_i$, (\ref{eqsgli}) is used. The calculations are 
for spherical wave propagation, appropriate for sources embedded in the the corona. The 
solid line uses an observing frequency, $f = 600$MHz, the dotted line uses $f = 327$MHz 
and the dashed line uses $f = 150$MHz.}
	\label{sphrat2}
\end{figure}

For sources embedded in the solar corona, spherical propagation effects are vital since the radiation is refracted along the radial direction. Since the assumption of isotropic scattering is justified in this situation (\S~2.2), the coherence lengths are computed using the formulation outlined in \S~2.1.2. An appropriate modification required for sources embedded in the solar corona will be discussed below.

For spherical wave propagation we need to use the effective baseline $s_{\rm eff} = sR/(R_1-R_0)$, where $R_1$ is the distance of the observer from the source and $R_0$ is the distance from which scattering is assumed to be effective. We consider $R_0$ to be equal to the fundamental plasma emission level; for 327MHz, with the wind-like density model $R_{0}$ is located at 0.0156 $R_{\odot}$ above the photosphere.

In this case, the line of sight from the source embedded in the solar corona to the observer (at the Earth) spans heliocentric distances $R$ ranging from the height of fundamental plasma emission (where $f_p(R) = f$) to 1 AU. One therefore needs to explicitly integrate equation (\ref{eqcsf}) along the line of sight with $R$ being the integration variable. This aspect is different from the plane wave case (\S~3.1.1).
Thus for the spherical wave propagation (\ref{eqcsf}) should be modified to,
\begin{eqnarray}
D_{\phi}(s) 
			= \frac{8\pi^2r_e^2\lambda^2}{2^{\alpha-2}(\alpha-2)}
			\Gamma\left(1-\frac{\alpha-2}{2}\right)
			\int_{R_0}^{R_1}\! \frac{C_N^2(R) l_i(R)^{\alpha-2}} {(1-f_p^2(R)/f^2)}
			\nonumber \\
			\times \left\lbrace_1F_1\left[-\frac{\alpha-2}{2},1, 
			-\left(\frac{sR/(R_1-R_0)}{l_i(R)}\right)^2 \right]-1  \right\rbrace \, 
   			\mathrm{d}R \,\, ,
\label{eqcsfsph}
\end{eqnarray}

\vskip1ex
\noindent and the corresponding asymptotic branches e.g. \citep{b14, b6, b39} by

\begin{eqnarray}
D_{\phi}(s) = \frac{4\pi^2 r_e^2 \lambda^2}{2^{\alpha-2}}
			\Gamma\left(1-\frac{\alpha-2}{2}\right) \left(\frac{s}{R_1-R_0}\right)^2 
			\nonumber \\
	 \times \int_{R_0}^{R_1}\! \frac{R^2 C_N^2(R)l_i(R)^{\alpha-4}}{(1-f_p^2(R)/f^2)} 
			\, \mathrm{d}R \,\,,
\label{eqsllisph}
\end{eqnarray} 
\vskip1ex
\noindent for $s_{\rm eff} \ll l_i$ and for $s_{\rm eff} \gg l_i$ by

\begin{eqnarray}
D_{\phi}(s) = \frac{8 \pi^2 r_e^2 \lambda^2}{2^{\alpha-2}(\alpha-2)}
			\frac{\Gamma\left(1-(\alpha-2)/2\right)}{\Gamma\left(1+(\alpha-2)/2\right)}
			\left(\frac{s}{R_1-R_0}\right)^{\alpha-2} \nonumber \\
	\times  \int_{R_0}^{R_1}\! \frac{R^{\alpha-2} C_N^2(R)}{(1-f_p^2(R)/f^2)} \, 
			\mathrm{d}R \,\, . 
\label{eqsglisph}
\end{eqnarray}

Furthermore, the lower limit of the integration (which is the fundamental plasma level) depends on the observing frequency; it therefore follows that the relative error in the coherence length also depends on the observing frequency. We use the structure function given by (\ref{eqcsfsph}) and (\ref{eqcohlength}) to find the coherence length predicted by the GSF. Similarly, we use (\ref{eqsllisph}), (\ref{eqsglisph}) and (\ref{eqcohlength}) to find the coherence lengths corresponding to the $s_{\rm eff} \ll l_{i}$ and $s_{\rm eff} \gg l_{i}$ branches respectively.

Figure~\ref{sphrat1} shows the relative error in the coherence length $s_{0}$ corresponding to the asymptotic branch $s_{\rm eff} \ll l_i$ for three different frequencies. Figure~\ref{sphrat1} shows that for 150MHz (dashed line), the $s_{\rm eff} 
\ll l_i$ branch is inadequate for $l_i \leq 10$ km, whereas for 327MHz (dotted line) and 600MHz (solid line) the $s_{\rm eff} \ll l_i$ branch is inadequate for $l_i \leq 20$ km and for $l_i \leq 60$ km respectively.

Figure~\ref{sphrat2} shows that, for an observing frequency of 150MHz (dashed line), the GSF predictions disagree with those of the $s_{\rm eff} \gg l_i$ branch for $l_i \geq 0.1$ km. On the other hand, for 327MHz (dotted line) and 600MHz (solid line) the GSF predictions disagree with those of the $s_{\rm eff} \gg l_i$ branch for $l_i \geq 0.4$ km and for $l_i \geq 1$km respectively.

To summarize, we find that the range of the inner scales for which the relative error is significant (i.e., the predictions of the GSF disagree with those of the asymptotic branches) is a weak function of the observing frequency. For observing frequencies ranging from 150 MHz to 600 MHz the GSF predictions disagree with (and are more accurate than) those of the asymptotic branches for 0.1 km $ \leq l_i \leq 60$ km.

\subsection{Effect of local flattening of the turbulence spectrum}

\begin{figure}
	\includegraphics[width=84mm]{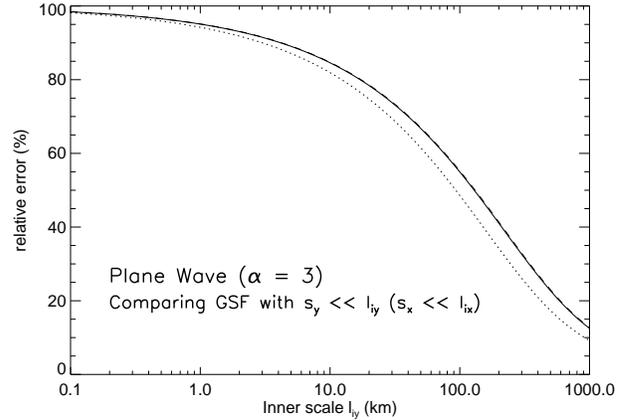}
	\caption{Relative error in the coherence length $s_{0y} \,\, (s_{0x})$ as a function of $l_{iy} \,\, (l_{ix})$ when the asymptotic branch $s_y \ll l_{iy} \,\, (s_x \ll l_{ix})$, (\ref{eqslliani}) is used with the power law index $(\alpha = 3)$. The calculations are for plane wave propagation through the corona and solar wind and for a representative solar elongation of $10R_{\odot}$. The dotted line uses the degree of anisotropy, $\rho = 1$, the solid line is for $\rho = 5$ and the dashed line uses, $\rho = 10$.}
	\label{nonkolmpln1}
\end{figure}

\begin{figure}
	\includegraphics[width=84mm]{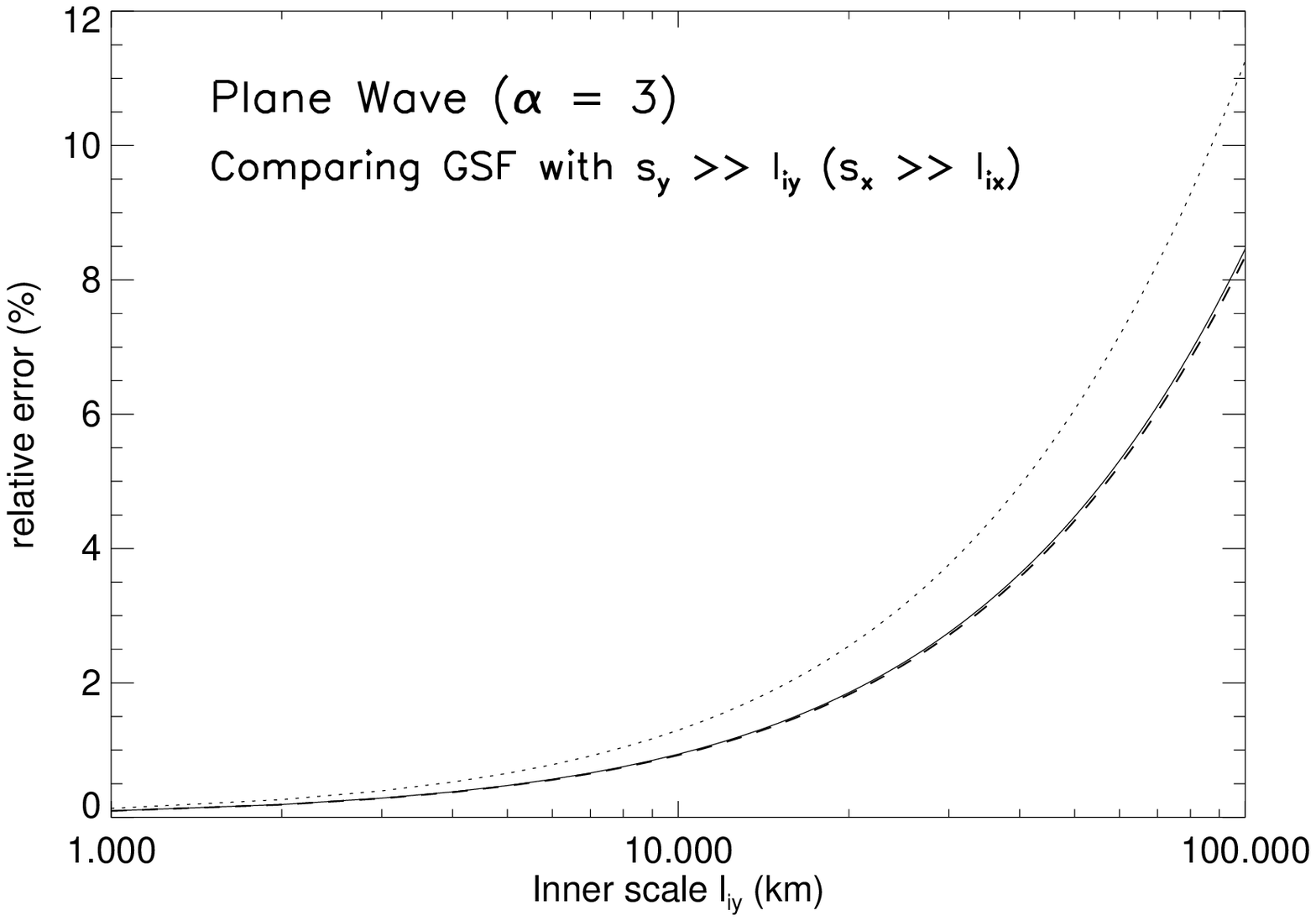}
	\caption{Relative error in the coherence length $s_{0y} \,\, (s_{0x})$ as a function of $l_{iy} \,\, (l_{ix})$ when the asymptotic branch $s_y \gg l_{iy} \,\, (s_x \gg l_{ix})$, (\ref{eqsgliani}) is used with the power law index ($\alpha = 3$). 
	The calculations are for plane wave propagation through the corona and solar wind and for a representative solar elongation of $10R_{\odot}$. The dotted line uses the degree of anisotropy, $\rho = 1$, the solid line is for $\rho = 5$ and the dashed line uses, $\rho = 10$.}
	\label{nonkolmpln2}
\end{figure}

\begin{figure}
	\includegraphics[width=84mm]{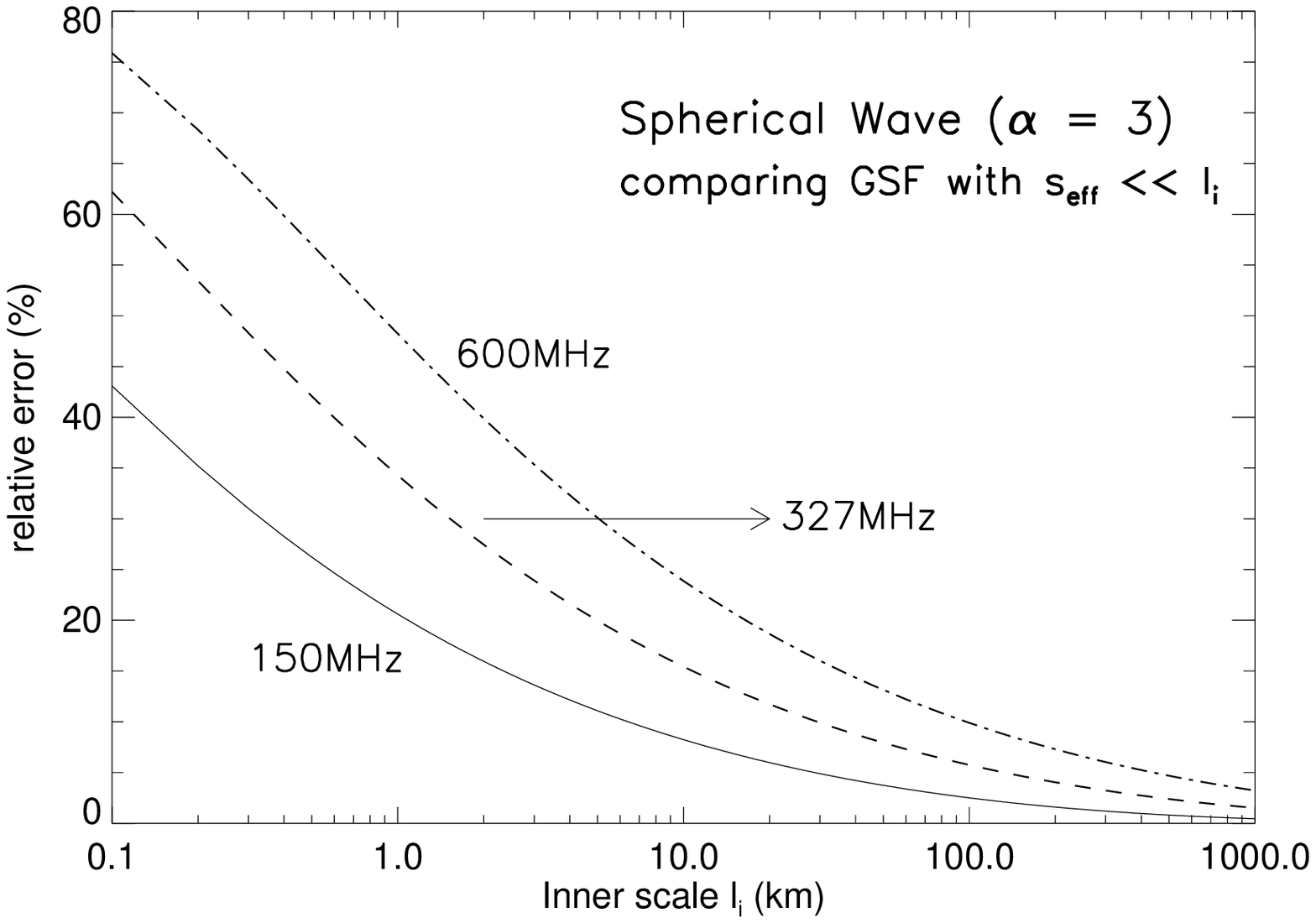}
	\caption{Relative error in the coherence length $s_{0}$ as a function of $l_i$ when 
the asymptotic branch $s_{\rm eff} \ll l_i$, (\ref{eqslli}) is used with the power law index $(\alpha = 3)$. The calculations are for spherical wave propagation, appropriate for sources embedded in the the corona. The solid line uses an observing frequency, $f = 150$MHz, the dashed line uses $f = 327$MHz and the dot-dashed line uses $f = 600$MHz.}
	\label{nonkolmsph1}
\end{figure}

\begin{figure}
	\includegraphics[width=84mm]{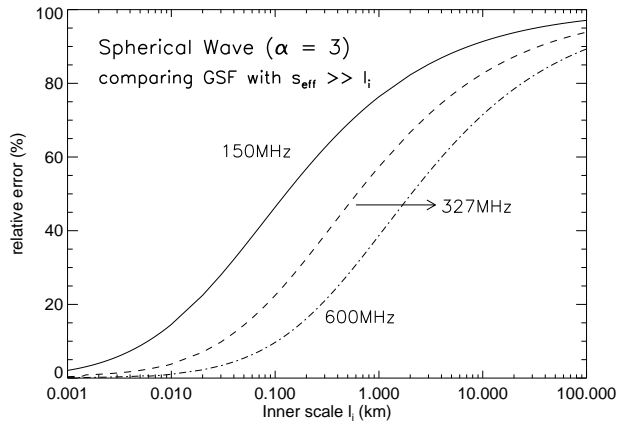}
	\caption{Relative error in the coherence length $s_{0}$ as a function of $l_i$ when 
the asymptotic branch $s_{\rm eff} \gg l_i$, (\ref{eqsgli}) is used with the power law index $(\alpha = 3)$. The calculations are for spherical wave propagation, appropriate for sources embedded in the the corona. The solid line uses an observing frequency, $f = 150$MHz, the dashed line uses $f = 327$MHz and the dot-dashed line uses $f = 600$MHz.}
	\label{nonkolmsph2}
\end{figure}

As mentioned earlier, there is some evidence for the flattening of the power spectrum of density turbulence between scales $\approx$ 100 km and the inner scale \citep{b15}. It is not clear how this feature evolves with heliocentric distance. Although our current formalism cannot accommodate two power laws and an exponential turnover, we follow 
\citet{b6} in adopting $\alpha = 3$ (instead of the Kolmogorov $\alpha = 11/3$) for the entire spectrum. As discussed in \S~2.2, the appropriate expression to use for $C_{N}^{2}$ is then (\ref{newcnsq}). 

With these modifications, Figures \ref{nonkolmpln1} and \ref{nonkolmpln2} show that for plane wave propagation (at an elongation of 10 $R_{\odot}$ and for $1 < \rho < 10$) the relative error in the coherence length $s_0$ is significant for 1 km $\leq l_i \leq 1000 $km. Furthermore, the region of disagreement is insensitive to the value of $\rho$ when $\rho \geq 5$.


For spherical wave propagation, Figures \ref{nonkolmsph1} and \ref{nonkolmsph2} show that, this range depends on the observing frequency. For 327 MHz we find that the GSF predictions for angular broadening observed at the Earth disagree with those of the asymptotic branches for $0.1$ km $\leq l_i \leq 100$ km. Thus, the range of $l_{i}$ over which the GSF and the asymptotic branch predictions disagree is larger for $\alpha = 3$ as compared to $\alpha = 11/3$.

\subsection{When are inner scale effects important?}

\begin{figure}
	\includegraphics[width=84mm]{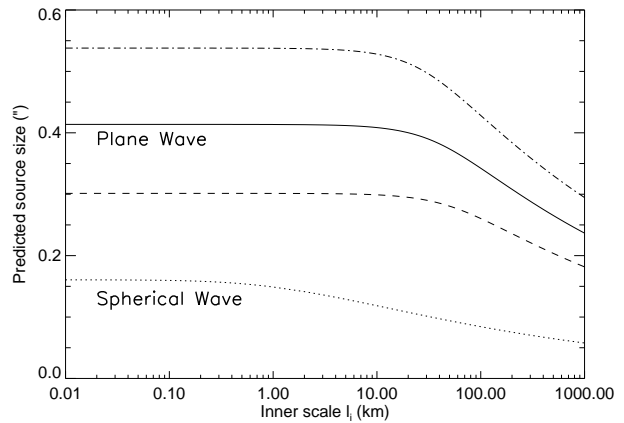}
	\caption{The predicted broadening $\theta_c$ as a function of the inner scale $l_i$ at an observing frequency of 327 MHz and Kolmogorov power law index $\alpha = 11/3$. The dashed, solid and dot-dashed lines are for $\theta_{cy}$ computed using GSF with plane wave propagation at a solar elongation of 10 $R_{\odot}$ and the degree of anisotropy $\rho = 1, 5$ and 10, respectively, and screen thickness $\Delta L = 0.5 
	R_{\odot}$, given by (\ref{eqcsfani}) while the dotted line is for $\theta_c$ computed using GSF with spherical wave propagation using (\ref{eqcsfsph}).}
\label{inscleff}
\end{figure}

We have established in Figures \ref{plnrat1} $-$ \ref{sphrat2} that it is essential to use the GSF for 4km $\leq l_i \leq 200$km for plane wave propagation and for 0.1km $\leq l_i \leq 60$km for spherical wave propagation. We next investigate the sensitivity of the predicted source size to $l_i$. For spherical wave propagation we calculate the scattering angle $\theta_c$ using (\ref{eqthetac}) with the GSF (\ref{eqcsfsph})and for plane wave propagation the scattering angle $\theta_{cy}$ is calculated using the GSF (\ref{eqcsfani}) and (\ref{aniangley}). We consider the inner scale $l_{i}$ (for spherical wave propagation) and $l_{iy}$  (for plane wave propagation) as a free parameter. 

Figure~\ref{inscleff} thus shows the extent of angular broadening for an ideal point source as a function of $l_i$, for plane wave and spherical wave propagation at an observing frequency of 327 MHz. For plane wave propagation these calculations are carried out at an elongation of 10 $R_{\odot}$ with a screen thickness $\Delta L = 0.5 R_{\odot}$ and $\rho = 1, 5$ and 10. It is clear from Figure \ref{inscleff} that for plane wave propagation, the extent of scatter broadening depends upon the value of the inner scale only for $l_{iy} \geq 10$ km. This gives the upper limit on the values of $l_i$ below which the results are independent of the inner scale. We find that this upper limit is a function of the degree of anisotropy, and it declines for larger values of $\rho$. For plane wave propagation, we find that inner scale effects are important only for heliocentric distances $\leq 20 R_{\odot}$. This result is consistent with our finding that the GSF can be approximated by the $s_y \gg l_{iy} \,\, (s_x \gg l_{ix})$ asymptotic branch for solar elongations $> 20 R_{\odot}$; this branch (\ref{eqsgliani}) does not involve the inner scale. On the other hand, for spherical wave propagation Figure \ref{inscleff} shows that the scatter broadening angle is sensitive to the inner scale for 
$l_i \geq 1$ km.

To summarize, for $f = 327$ MHz, inner scale effects can generally be considered to be important (in the sense that the source size using the GSF is sensitive to the actual value of the inner scale) if $l_{i} \geq$ few hundred meters to a few km. We have carried out similar calculations for $f = 1500$MHz; for this frequency, we find that the source size is sensitive to the inner scale if $l_i \geq$ a few km to 100 km.

\subsection{Inner scale models}

We have established with the preceding calculations the range of inner scale values for which the GSF needs to be used (\S~3.1.1 and \S~3.1.2), maintaining the inner scale as a free parameter in our investigations so far. We now evaluate the inner scale in the corona and the solar wind using three different physical prescriptions. The first prescription is one where the inner scale arises from proton cyclotron damping by MHD waves \citep{b15, b47, b7}:

\begin{equation}
l_i(R) = 684\times n_e(R)^{-1/2}   \,\,\,\,\,\,\,\,\,     {\rm km} \,\, ,
\label{ch89}
\end{equation}

\noindent where $n_e$ is the electron number density in cm$^{-3}$. For the second prescription we identify the inner scale with the proton gyroradius \citep{b5, b1}: 

\begin{equation}
l_{i}(R) =  1.02\times 10^2 \mu^{1/2} T_i^{1/2} B(R)^{-1}  \,\,\,\,\,\,\, {\rm cm} \,\, ,
\label{protonscale}
\end{equation}

\noindent where $\mu  \,\, (\equiv m_p/m_e)$ is the proton to electron mass ratio, $T_i$ is the proton temperature in eV and B is the Parker spiral magnetic field in the ecliptic plane 
\citep{b43}. However, recent work seems to suggest that the dissipation could occur at scales as small as the electron gyroradius \citep{b1, b34}. The third prescription we consider is thus that the inner scale is the electron gyroradius $\rho_e$, given by:

\begin{equation}
l_{i}(R) = 2.38\times T_e^{1/2} B(R)^{-1}  \,\,\,\,\,\,\,\,\,   {\rm cm} \,\, ,
\label{electronscale}
\end{equation}

\noindent where $T_e$ is the electron temperature in eV.

\begin{figure}
   	\includegraphics[width=84mm]{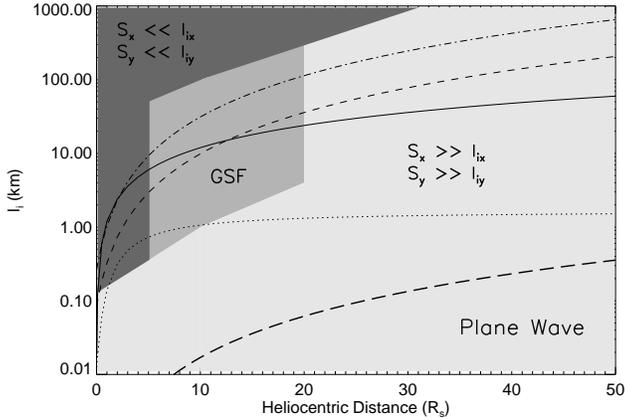}
	\caption{The inner scale $l_{iy} \,\, (l_{ix})$ in km as a function of heliocentric distance in radius of Sun ($R_s$), for plane wave propagation. The dashed and dot-dashed lines show the proton gyroradius (\ref{protonscale}) using proton temperatures of $10^5$ and $10^6$ K respectively. The solid and dotted lines show the inner scale governed by proton cyclotron damping (\ref{ch89}) using the wind-like density model (\ref{wind}) and the fourfold Newkirk density model respectively. The thick dashed line shows the electron gyroradius (\ref{electronscale}) using an electron temperature of $10^5$K. The light grey region denotes the range of (distant) source elongations and inner scale values for which the GSF yields predictions that are substantially more accurate than those of the asymptotic branches.}
\label{inscl}
\end{figure}

\begin{figure}
   	\includegraphics[width=84mm]{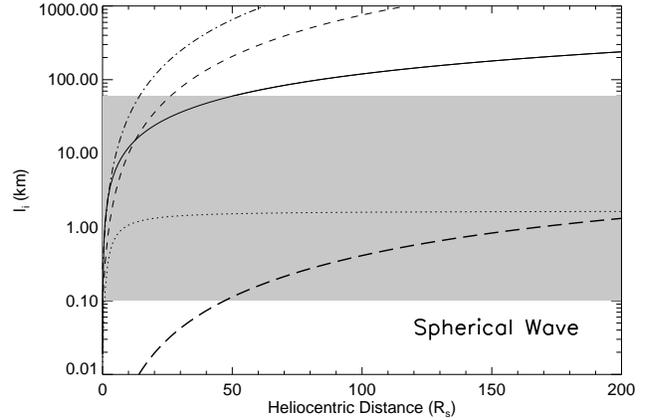}
	\caption{The inner scale $l_i$ in km as a function of heliocentric distance in radius of Sun ($R_s$), for spherical wave propagation. The dashed and dot-dashed lines show the proton gyroradius (\ref{protonscale}) using proton temperatures of $10^5$ and $10^6$ K respectively. The solid and dotted lines show the inner scale governed by proton cyclotron damping (\ref{ch89}) using the wind-like density model (\ref{wind}) and the fourfold Newkirk density model respectively. The thick dashed line shows the electron gyroradius (\ref{electronscale}) using an electron temperature of $10^5$K. The light grey region denotes the range of (distant) source elongations and inner scale values for which the GSF yields predictions that are substantially more accurate than those of the asymptotic branches.}
\label{inscl_sph}
\end{figure}

\indent Figure~\ref{inscl} shows the inner scale obtained using these three prescriptions as a function of heliocentric distance. It is useful to compare the predictions of the inner scale models with the range of inner scales for which we claim that the GSF needs to be used. For plane wave propagation, a distant cosmic source is located at a given solar elongation (which we take to be the same as the heliocentric distance for the purposes of this discussion) behind the solar wind scattering screen. At this heliocentric distance, the angular broadening prediction using the GSF is more accurate than that of either of the asymptotic branches if 0.3km $\leq l_{i} \leq$ 300 km (\S~3.1.1). The light grey region in Figure \ref{inscl} denotes this region; it indicates the range of inner scales for which the GSF predictions are more accurate than those of the asymptotic branches for plane wave propagation for distant cosmic sources located at solar elongations between $5 R_{\odot}$ and $20 R_{\odot}$ and having anisotropies in the range of 1$-$10. The $s \ll l_{i}$ asymptotic branch is adequate for elongations $< 5 R_{\odot}$ (dark grey region in Figure \ref{inscl}) , while the  $s \gg l_{i}$ asymptotic branch is adequate for elongations $> 20 R_{\odot}$. To summarize, Figure~\ref{inscl} reveals that, for distant cosmic sources (for which plane wave propagation is appropriate) located at solar elongations between $5 R_{\odot}$ and $20 R_{\odot}$ and anisotropies
$1 \leq \rho \leq 10$, the GSF would need to be used if the inner scale is the proton gyroradius or is due to proton cyclotron resonance.
These results are summarized in Table~\ref{table_pln}

\begin{table*}
 \centering
 \begin{minipage}{140mm}
  \caption{Plane wave propagation with degree of anisotropy $\rho$ between 1$-$10 : when does the GSF need to be used? (GSF required for 0.3km $\leq l_i \leq 300$km).}
  \label{table_pln}
  \begin{tabular}{@{}clrlrcc@{}}
  \hline
 	 & 	\multicolumn{6}{c}{Inner scale model (km)} \\
  {Solar Elongation}	 &	\multicolumn{2}{c}{Proton Gyroradius} & \multicolumn{2}{c}{proton cyclotron 
  	 					damping} & \multicolumn{2}{c}{electron Gyroradius} \\
  $R_{\odot}$	 & 	$T_p = 10^5$ K & $T_p = 10^6$ K & wind like & 4* Newkirk & $T_e = 10^5$ K 
  	 							         & $T_e = 10^6$ K\\
  	 &  &  & density model & density model &  &  \\
 \hline
 $< 5$ & $s \ll l_i$ is valid & $s \ll l_i$ is valid & $s \ll l_i$ is valid & $s \ll l_i$ is valid &  &  \\
 		&	&	&	&	& $s \gg l_i$ is &	$s \gg l_i$ is \\
 $5< R < 20$  & GSF required & GSF required & GSF required & GSF required & valid & valid \\
 							   
  &   &  &  &  & throughout & throughout \\
 							
 $> 20$ & $s \gg l_i$ is valid & $s \gg l_i$ is valid & $s \gg l_i$ is valid &  $s \gg l_i$ is valid & \\

 										  	   
 \hline
\end{tabular}
\end{minipage}
\end{table*}


\begin{table*}
 \centering
 \begin{minipage}{140mm}
  \caption{Spherical wave propagation: when does the GSF need to be used?.}
  \label{table_sph}
  \begin{tabular}{@{}llrrrcc@{}}
  \hline
 	 & 	\multicolumn{6}{c}{Inner scale model (km)} \\
  	 &	\multicolumn{2}{c}{Proton Gyroradius} & \multicolumn{2}{c}{proton cyclotron 
  	 	damping} & \multicolumn{2}{c}{electron Gyroradius} \\
  	 & 	$T_p = 10^5$ K & $T_p = 10^6$ K & wind like & 4* Newkirk & $T_e = 10^5$ K 
  	 								 & $T_e = 10^6$ K\\
  	 &  &  & density model & density model &  &  \\
 \hline
 Is GSF required ? &  &  &  &  & No, & No,  \\
 (GSF required if  & Yes & Yes & Yes & Yes & GSF required  
 							   & GSF required \\
 0.1km $\leq l_i \leq 60$km) &  &  &  &  & only from 70 $R_{\odot}$ 
 					 & only from 70 $R_{\odot}$ \\
  & & & & &	to the Earth & to the Earth \\
 										  	   
 \hline
\end{tabular}
\end{minipage}
\end{table*}

We carry out a similar exercise for spherical wave propagation. In this situation, the source is embedded in the solar corona and the observer is at the Earth, looking at the source through the turbulent medium. Figure~\ref{inscl_sph} shows the inner scale obtained using the three prescriptions described above as a function of heliocentric distance. The linestyles are the same as those used in Figure~\ref{inscl}. As explained in \S~3.1.2, for spherical wave propagation at observing frequencies ranging from 150 MHz to 600 MHz the predictions of the GSF are more accurate than those of the the asymptotic branches for 0.1km $\leq l_{i} \leq$ 60km. This region is represented by a grey band in Figure~\ref{inscl_sph}. It is well known that most of the scattering takes place well within $30 R_{\odot}$ 
\citep{b39}. We can claim that the angular broadening estimates using the GSF will be more accurate than those of the asymptotic branches if the grey band in Figure~\ref{inscl_sph} encloses the inner scale predicted by a specific model for heliocentric distances $\leq$ 30 $R_{\odot}$. An examination of Figure~\ref{inscl_sph} show that this is the case (i.e., the GSF needs to be used for accurate broadening estimates) if the inner scale is governed by proton-cyclotron damping or is given by proton gyroradius. If, on the other hand, the inner scale is the electron gyroradius with $T_e = 10^5$K, the inner scale values predicted by this model overlap the grey band in Figure~\ref{inscl_sph} only for heliocentric distances $\geq$ 30 $R_{\odot}$. For $T_e = 10^6$K, this is true for heliocentric distances $\geq$ 50 $R_{\odot}$. Thus, if the inner scale is given by the electron gyroradius, we cannot claim that the GSF predictions will be substantially more accurate than the predictions of the asymptotic branches. These results are summarized in Table~\ref{table_sph}.

\section{Summary and conclusions}

The amplitude of MHD turbulence in the extended solar corona and solar wind, especially near the inner (dissipation) scale, is a subject that is of considerable interest in a variety of applications. We investigate it using predictions for the angular broadening of radio sources. Most estimates of angular broadening due to refraction and scattering by density turbulence employ approximations to the structure function that hold for situations where the interferometer spacing is either much larger than, or much smaller than the inner scale $l_{i}$. We use a general structure function (GSF) that does not rely on these approximations. We consider both plane wave propagation, which is appropriate for distant cosmic sources observed against the background of the solar wind, and spherical wave propagation, which is appropriate for sources embedded in the solar corona. For plane wave propagation we consider an anisotropic density turbulence spectrum comprising a Kolmogorov power law ($\alpha = 11/3$) spectrum multiplied by an exponential turnover at the inner scale. For spherical wave propagation, isotropic scattering is a well justified assumption. We demonstrate that angular broadening predictions using the general structure function agree with those obtained using the appropriate asymptotic expressions in the limits $s \ll l_{i}$ and $s \gg l_{i}$. For plane wave propagation, for sources observed at elongations between 5 and 20 $R_{\odot}$ and with the degree of anisotropy $1 \leq \rho \leq 10$, we find that the GSF needs to be used for 4 km $\leq l_{ix}, \, l_{iy} \leq$ 200 km. These results are independent of observing frequency as well as the amplitude of the density turbulence ($C_N^2$), and only weakly dependent on the degree of anisotropy ($\rho$). For spherical wave propagation, however, the results are found to be weakly dependent on the observing frequency. For observing frequencies ranging from 150 MHz to 600 MHz, the predictions of the GSF are more accurate than those of the asymptotic branches if 0.1km $\leq l_i \leq$ 60 km. If the spectrum is taken to be flatter ($\alpha = 3$), then the range of $l_{i}$ for which the GSF predictions disagree with those of the asymptotic branches is larger. Importantly, the range over which the GSF predictions are substantially more accurate than those of the asymptotic approximations for plane wave propagation (light grey band in Figure~\ref{inscl}) is well within the expected values of the inner scale for several different models (proton cyclotron damping and the proton gyroradius). For plane wave propagation, we find that angular broadening predictions using the GSF are sensitive to the value of the inner scale for distant cosmic source located at elongations $\leq 20 R_{\odot}$. For spherical wave propagation, which is applicable when a source embedded in the solar corona is viewed at the Earth, the GSF is more accurate if the inner scale is due to proton cyclotron damping or is given by the proton gyroradius.

Using the GSF with spherical wave propagation to calculate the predicted extent of broadening of an ideal point source, we find that the extent of angular broadening is sensitive to the value of $l_{i}$ (in other words, inner scale effects are significant) if $l_{i} \geq$ a few to a few tens of km for $f = 327$ MHz. For an observing frequency of 1500 MHz, inner scale effects are important if $l_{i} \geq$ a few to 100 km. 

The rate at which energy in solar wind turbulence damps on ions is an important question cutting across sub-disciplines. While some progress has been made in this regard, its still not clear if there is enough energy in the cascade near the dissipation scale for direct perpendicular heating \citep{b17}. This question can be addressed via accurate estimates of the amplitude of density turbulence ($C_{N}^{2}$). Observations of angular broadening of radio sources are typically reliable means of constraining $C_{N}^{2}$. Recent conclusions regarding the magnitude of density fluctuations (relative to the background density) in the heliosphere \citep{b51} are also expected to help in constraining $C_{N}^{2}$. However, such estimates have traditionally been made using expressions for the structure function that are only valid in limits where the interferometric baseline used for observing are either $\gg$ or $\ll$ the dissipation scale. We have used the general structure function and quantified the errors arising from the use of these approximations. Our results underline the necessity of using the GSF for quantitative estimates of angular broadening.

\section*{Acknowledgments}

MI acknowledges support via a PhD studentship from the Indian Institute of Science Education and Research, Pune. MI and PS acknowledge support from the CAWSES - II program administered by the Indian Space Research Organization. IHC acknowledges the Australian Research Council for support. {\bf We thank the referee for insightful comments and suggestions that have helped improve the 
paper.}

\section*{Appendix : Pseudocode for GSF}

In this appendix we present psuedocodes in Mathematica and python to implement the confluent hypergeometric function
which is the main feature of the GSF.\\ 

Mathematica : The Confluent Hypergeometric function of the first kind is implemented 
in mathematica as Hypergeometric1F1[a,b,z]. We use Numerical Root Finding packages e.g. FindRoot (which implements the Newton-Rhapson method), to obtain the root of the equation \, :\,
$D_{\phi}(s_0) - 1.0 = 0$\\

\emph{Mathematica Pseudocode :} \\

For $b > 0$ \\
\indent hyp[$a_{-},b_{-},z_{-}$] := Hypergeometric1F1[a,b,z] \\

define $D_{\phi}(s)$\\

FindRoot$[D_{\phi}(s) - 1.0 == 0.0, \,\, \{s, s_0\}]$ \\
					
Python : Similar packages are available in python, which is an open source programming langauge. The Confluent 
Hypergeometric function of the first kind is available under mpmath package as hyp1f1(a,b,z) and findroot from 
mpmath impliments the secant method for root finding by default \\

\emph{Python Pseudocode :} \\

from mpmath import * \\
\indent def hyp(a,b,z):\\
\indent	\qquad return hyp1f1(a,b,z) \\

define $D_{\phi}(s)$\\

mp.dps = 30; mp.pretty = True \\
\indent rt = findroot $(D_{\phi}(s)-1, \,\,\, s_0) $ \\

\label{lastpage}

\end{document}